\documentclass[prl,superscriptaddress,twocolumn,showpacs]{revtex4}
\usepackage[ansinew]{inputenc}
\usepackage{graphicx}
\usepackage{amsmath}
\usepackage{amsthm}
\usepackage{bm}
\usepackage{layout}
\usepackage{float}
\usepackage{amsfonts}
\usepackage{amssymb}
\begin{document}

\title{Experimentally friendly geometrical criteria for entanglement}

\author{Piotr Badzi{\c{a}}g}
\affiliation{Alba Nova Fysikum, University of Stockholm, S--106~91, Sweden}
\affiliation{Institute of Theoretical Physics and Astrophysics, University
of Gda\'nsk, ul. Wita Stwosza 57, PL-80-952 Gda\'nsk, Poland}
\affiliation{National Quantum Information Centre of Gda\'nsk, ul. W. Andersa 27, PL-81-824 Sopot, Poland}

\author{{\v C}aslav Brukner}
\affiliation{Institute for Quantum Optics and Quantum Information,
Austrian Academy of Sciences, Boltzmanngasse 3, A-1090 Vienna, Austria}
\affiliation{Faculty of Physics, University of Vienna, Boltzmanngasse 5, A-1090 Vienna, Austria}

\author{Wies{\l}aw Laskowski}
\affiliation{Institute of Theoretical Physics and Astrophysics, University
of Gda\'nsk, ul. Wita Stwosza 57, PL-80-952 Gda\'nsk, Poland}

\author{Tomasz Paterek}
\affiliation{Institute for Quantum Optics and Quantum Information,
Austrian Academy of Sciences, Boltzmanngasse 3, A-1090 Vienna, Austria}

\author{Marek \.Zukowski}
\affiliation{Institute of Theoretical Physics and Astrophysics, University
of Gda\'nsk, ul. Wita Stwosza 57, PL-80-952 Gda\'nsk, Poland}

\pacs{03.65.Ud}

\begin{abstract}
A simple geometrical criterion 
gives experimentally friendly sufficient
conditions for entanglement.  Its  generalization gives
a necessary and sufficient condition. It is
linked with a family of entanglement identifiers, which is strictly
richer than the family of entanglement witnesses. 
\end{abstract}

\date{\today}

\maketitle

Entanglement is one of the basic features of quantum physics
and it is a resource for quantum information science
\cite{NIELSENCHUANG}. Thus,
 detection of entanglement belongs to the mainstream of 
this field \cite{HORODECCY_RMP}. Today, the most widely used  and
experimentally feasible detectors of this resource  are 
entanglement witnesses \cite{ENT_WITNESS}. They are linked
with positive but not completely positive maps \cite{JAMIOLKOWSKI},
which are the most universal entanglement identifiers. 

We present an alternative approach to
entanglement detection. It is rooted in an elementary geometrical fact: 
if a scalar product of two real vectors $\vec{s}$ and $\vec{e}$ satisfies $\vec{s} \cdot \vec{e}< \vec e \cdot \vec e$, then $\vec{s}\neq\vec{e}$. 
This fact was used in, e.g., \cite{ZUKOWSKI} to derive a powerful series of Bell inequalities, 
and in \cite{LASKOWSKI-ZUK} it led to 
sufficient condition for entanglement. 
Here, it inspires a new family of entanglement identifiers, which are naturally expressed in terms of the correlation functions \cite{CASAL}, easily determined by local measurements. 
This makes them friendly to experiments. The family of our identifiers is richer than the family of the entanglement witnesses and leads to a necessary and sufficient criterion for entanglement. 

The bulk of our \emph{presentation} uses systems of many spin-$\frac{1}{2}$ particles (qubits), but the method is applicable to composite systems of arbitrary dimensions. For that in our formulae, one needs to substitute Pauli operators by their Gell-Mann-type generalizations. This  allows a complete separability analysis of a multi-partite state, and will be illustrated by an example. 
Even if the underlying system consists of many qubits, analysis of the so-called $k$-separability ($k<N$) requires identification of entanglement in the system partitioned into $k$ parts only \cite{DurCirac00}. Clearly, at least one part will contain two or more qubits,
and can be considered as a multi-level system. 

An $N$-qubit density matrix can be put as follows:
\begin{equation}
\rho = \frac{1}{2^N} \sum_{\mu_1,...,\mu_N=0}^3 T_{\mu_1...\mu_N}
\sigma_{\mu_1} \otimes ... \otimes \sigma_{\mu_N},
\label{STATE}
\end{equation}
where $\sigma_{\mu_n} \in \{\openone,\sigma_x,\sigma_y,\sigma_z\}$ is the
$\mu_n$'th local Pauli operator of the $n$th party ($\sigma_0= \openone$) and $T_{\mu_1...\mu_N} \in [-1,1]$ are the
components of the (real) extended correlation tensor $\hat T$. 
They are the expectation values 
$T_{\mu_1...\mu_N} = \mbox{Tr}[\rho (\sigma_{\mu_1} \otimes ... \otimes
\sigma_{\mu_N})]$.
The extended tensors are elements of a real vector
space with the scalar product given by
\begin{equation}
\label{ScalProd}
(\hat X,\hat Y) = \sum_{\mu_1,...,\mu_N=0}^3 X_{\mu_1...\mu_N} Y_{\mu_1...\mu_N}.
\end{equation}

A state $\rho$ is separable if it can be put as a
convex combination
of product states, i.e.,
\begin{equation}
\rho_{\mathrm{sep}} = \sum_{i} p_i \rho_i^{(1)} \otimes ... \otimes
\rho_i^{(N)},
\label{SEP}
\end{equation}
with $p_i \ge 0$ for all $i$, and $\sum_i p_i = 1$.
Such a state
is specified by a separable extended tensor $\hat T^{\mathrm{sep}} =
\sum_{i} p_i \hat T^{\mathrm{prod}}_i$,
where $\hat T^{\mathrm{prod}}_i = \hat T^{(1)}_i \otimes ... \otimes \hat
T^{(N)}_i$
and each $\hat T^{(k)}_i$ describes one qubit state.
Thus, definition (\ref{SEP}) implies the following
simple criterion.
If state $\rho$, endowed with an extended correlation tensor $\hat{T}$, is
separable, then there is an (extended) correlation tensor of a pure
product state, $\hat T^{\mathrm{prod}}$, such that
$
(\hat T,\hat T^{\mathrm{prod}}) \ge (\hat T,\hat T).
\label{IMPLY1}
$
Indeed, assume that $(\hat T,\hat T^{\mathrm{prod}}) < (\hat T,\hat T)$
for all product states.
Due to separability of $\hat T$, it implies that 
$$(\hat T,\hat T) = \sum_{i} p_i (\hat T,\hat T_i^{\mathrm{prod}}) <
\sum_{i} p_i (\hat T,\hat T) = (\hat T, \hat T),$$
which is a contradiction.   
In other words,
\begin{equation}
\textrm{if} \hspace{5mm} \max_{\hat T^{\mathrm{prod}}} \textrm{ } (\hat T,\hat T^{\mathrm{prod}}) <
(\hat T,\hat T),
\label{SIMPLE_WITNESS}
\end{equation}
then $\rho$ is entangled.

A simple entanglement identifier is obtained if in the space of correlation tensors
one introduces an improper scalar product with the summation indices $\mu_n$ in
(\ref{ScalProd}) running through the values  $j_n = 1,2,3$ only (sometimes referred to as $x,y,z$), i.e., 
\begin{equation}
\label{ScalProd1}
(\hat{X}_N,\hat{Y}_N) = \sum_{j_1,...,j_N=1}^3 X_{j_1...j_N} Y_{j_1...j_N}.
\end{equation}
The maximum of the left-hand side of
(\ref{SIMPLE_WITNESS}) is now given by 
the highest generalized Schmidt coefficient \cite{MULTISCHMIDT} of tensor $\hat{T}_N$, denoted here as $T_N^{\max}$. 
This is the maximal value of the $N$-qubit correlation function,
$T_N^{\max} = \max_{\vec m_1 \otimes ... \otimes \vec m_N} (\hat T,\vec m_1 \otimes ... \otimes \vec m_N)$
where $\vec m_n = (T^{(n)}_x,T^{(n)}_y,T^{(n)}_z)$ is a three-dimensional unit vector
describing a pure state of the $n$th party.
Therefore, 
\begin{equation}
\mathcal{E} = \frac{||\hat{T}_N||^2}{T_N^{\max}},
\label{SIMPLE_E}
\end{equation}
with $||\hat{T}_N||^2 = (\hat{T}_N,\hat{T}_N)$,
is a simple entanglement identifier.
If $\mathcal{E} > 1$, the state is non-separable.

A similar result is obtained when the summation indices are restricted to
$j_n = 1,2$. In this case, however, $T_N^{\max}$ refers to a maximization
restricted to two-dimensional sections of the correlation tensor.

If one finds that $||\hat{T}_N||^2 > 1$, one can immediately conclude that
the measured state is entangled because
 $T_N^{\max} \le 1$. Entanglement identification is then reduced
to measurements of orthogonal components $T_{j_1...j_N}$ of the
correlation tensor and summing up their squares until 
$\sum_{j_1,...,j_N} T_{j_1...j_N}^2$ exceeds unity. 
In some cases few measurements may suffice. 
Take for example the Greenberger-Horne-Zeilinger (GHZ) state \cite{GHZ}:
\begin{equation}
| \mathrm{GHZ}_N \rangle = \frac{1}{\sqrt{2}} \left( |z+\rangle_1...|z+\rangle_N + | z- \rangle_1...|z- \rangle_N \right),
\end{equation}
where $| z \pm \rangle$ are the eigenstates of the $\sigma_z$ operator.
For indices $x$ or $y$, this state has $2^{N-1}$ components  of the correlation tensor equal  to $ \pm 1$.
Measurement of any \emph{two} of them is sufficient to detect
entanglement.
Likewise, two measurements suffice to detect entanglement in any of the
graph states \cite{GRAPH} definable by $N$-qubit correlations.

Although very simple, the entanglement identifier
of Eq.~(\ref{SIMPLE_E}) is  optimal for some cases.
Consider Werner states, 
e.g. a mixtures of a singlet $| \psi^- \rangle$
with white noise, with the respective weights $p$ and $1-p$.
The extended correlation tensor of this state is diagonal, with the entries
$(1,-p,-p,-p)$.
Thus,  $T_2^{\max} = p$,  while one has $||\hat{T}_2||^2 = 3 p^2$.
Thus, $\mathcal{E} > 1$ for all entangled states of the family, i.e.
for $p > \frac{1}{3}$.
The new tool identifies entanglement also if the singlet state is replaced
by any other maximally entangled state. This distinguishes our identifiers from linear witnesses.
There is no single linear witness, which detects entanglement of all Bell states.

\emph{A higher dimensional example: qutrits.}
An arbitrary state of a single qutrit can be parameterized by an $8$-dimensional real vector,
whose components are the mean values of Gell-Mann operators.
Pure states correspond to normalized vectors $\vec n$
(we use the same normalization factors as in Ref.~\cite{WODKIEWICZ}).
The admissible vectors $\vec n$ obey an additional condition, see e.g.~\cite{WODKIEWICZ}.
A state of two qutrits can be expressed
in the operator basis made of tensor products of  Gell-Mann operators (including the unit operators).
Consider a mixture of maximally entangled state of two qutrits
$| \Psi \rangle = \frac{1}{\sqrt{3}}(|11 \rangle + |22 \rangle + | 33 \rangle)$
and white noise, with respective weights $p$ and $1-p$.
The (not extended) correlation tensor of this state is diagonal:
$\hat T_2 = \frac{p}{2} \mathrm{diag}[1,-1,1,1-1,1-1,1]$ and $||\hat T_2||^2=2p^2$.
The maximization in our criterion is over all normalized vectors $\vec n_A$ and $\vec n_B$. 
It gives $\max_{\vec n_A, \vec n_B} (\hat T_2,\vec n_A \otimes \vec n_B)=\frac{p}{2}$.
This cannot be smaller than the maximum over the admissible $\vec n$'s only.
Thus,
the state is entangled for $p > \frac{1}{4}$.
The same value is reported in, e.g., Ref. \cite{WODKIEWICZ}.
The fact that this can be obtained ignoring the condition for admissible $\vec n$'s is a surprising bonus.

{\em Generalized Werner states of $N$ qubits.} Consider
mixtures of a GHZ state with the white noise:
\begin{equation}
\rho(p) = p | \mathrm{GHZ}_N \rangle \langle \mathrm{GHZ}_N | + (1-p)
\frac{1}{2^N} \openone. 
\label{ONE}
\end{equation}
Since white noise exhibits no correlations, the
components $T_{j_1...j_N}$ of the correlation tensor of $\rho(p)$
are related to the components $T^{\mathrm{GHZ}}_{j_1...j_N}$ of the GHZ
state by  $T_{j_1...j_N}~=~p~T^{\mathrm{GHZ}}_{j_1...j_N}$.
Again, $T^{\max} = p$.
Applying condition~(\ref{SIMPLE_E}) with the sums over $j_n=1,2,3$
one finds that $\mathcal{E}>1$ for the admixture parameter 
$p > 1/(2^{N-1}+ \frac{1+(-1)^N}{2})$.
The same critical value for $N$ even was found by Pittenger and Rubin~\cite{PR},
who used the PPT criterion.
However, for $N$ odd their result is still $p > 1/(2^{N-1}+ 1)$
whereas in our case the term $\frac{1+(-1)^N}{2}$ vanishes
and our criterion in its simplest form is weaker than PPT.

The reason why condition (\ref{SIMPLE_E})
is not as efficient as PPT is that
for odd $N$ the GHZ states have no $T_{z...z}$ correlations.
Nevertheless, they have additional correlations between local $z$
measurements on even numbers of particles. These correlations are
described by their corresponding $2^{N-1}-1$ components of the extended correlation tensor. 
Our simplest criterion does not utilize these correlations. If one
attempts to include them
by using condition (\ref{SIMPLE_WITNESS}) with indices $\mu_n=0,1,2,3$
the situation gets even worse.

\emph{Generalized scalar product}.
The last example indicates that taking into account more correlations in the criterion does not guarantee better entanglement detection. For a success, 
one needs a proper combination of the correlations. To identify it, one may consider generalized scalar products, defined via a positive
semi-definite metric $G$:
\begin{equation}
(\hat X, \hat Y)_G = \sum_{\substack{\mu_1,...,\mu_N, \\ \nu_1,...,\nu_N=0}}^3
X_{\mu_1...\mu_N} G_{\mu_1...\mu_N,\nu_1...\nu_N} Y_{\nu_1...\nu_N}.
\end{equation}
If one can find a metric for which 
\begin{equation}
\max_{\hat T^{\mathrm{prod}}} \textrm{ } (\hat T,\hat T^{\mathrm{prod}})_G
< (\hat T,\hat T)_G,
\label{WEIGHTED_CONDITION}
\end{equation}
then the state $\rho$ described by its (extended) correlation tensor $\hat T$ is
 entangled.
 
This  criterion is very powerful and 
often it is easy to apply: a suitable metric  can be guessed from the structure of the
correlation tensor of the state in question.
Later on we will prove that criterion (\ref{WEIGHTED_CONDITION}) is also necessary for a state to
be entangled.

\emph{Generalized Werner states for odd $N$}.
To illustrate criterion
(\ref{WEIGHTED_CONDITION}), let us return to the generalized Werner
states for an odd number of qubits. 
Consider a diagonal metric $G_{\mu_1...\mu_N,\nu_1...\nu_N} =
G_{\mu_1...\mu_N} \delta_{\mu_1...\mu_N,\nu_1...\nu_N}$, with $G_{0...0}=0$, all $G_{j_1...j_N}=1$, all $G_{\mu_1...\mu_N}$ with
at least one $\mu_n=0$ equal to $\omega=1/(2^{N-1}-1)$.
This makes the left-hand side of condition (\ref{WEIGHTED_CONDITION}) equal to
$p(2^{N-1}-1)\omega = p$. This is seen directy once one writes down the `spatial' components of vectors defining single qubit states, $\hat T^{(n)}$, $n=1,..., N$, in the spherical coordinates. 
The optimal choice is to put all the local vectors along $z$ directions.
The right-hand side of (\ref{WEIGHTED_CONDITION}) is given by $p^2
(2^{N-1} + (2^{N-1}-1)\omega)=p^2(2^{N-1}+1)$.
Thus, the condition reveals entanglement of the generalized Werner states
for $p > 1/(2^{N-1}+1)$, 
exactly as given by the PPT criterion.

\emph{Colored noise}.
Consider, e.g., a two-qubit state
\begin{equation}
\rho_C(p) = p | \psi^- \rangle \langle \psi^- | + (1-p) |z+ \rangle
\langle z+ | \otimes |z+ \rangle \langle z+ |.
\label{COLORED_NOISE}
\end{equation}
Its correlation tensor has six non-vanishing elements: $T_{00} = 1$,
$T_{xx} = T_{yy} = -p$, $T_{zz} = 1-2p$, and $T_{z0} = T_{0z} = 1-p$.
The whole range of $p$, for which $\rho_C(p)$ is entangled,
can be found using a diagonal metric with the following non-zero elements:
$G_{11} = G_{22} = 1$ and $G_{03} = G_{30} = p$.
To find maximum of (\ref{WEIGHTED_CONDITION}), we again use the spherical coordinates
for the spatial elements of
the vectors describing the single qubit pure states: 
$T^{(n)}_x = \sin \theta_n \cos \varphi_n$, $T^{(n)}_y = \sin \theta_n
\sin \varphi_n$, and $T^{(n)}_z = \cos \theta_n$.
This gives for the left-hand side
$L = \max_{\theta_1,\theta_2,\varphi_1,\varphi_2}[(\cos\theta_1 
\cos\theta_2 - \sin\theta_1 \sin\theta_2 \cos(\varphi_1 - \varphi_2))p -
(\cos\theta_1 + \cos\theta_2)p^2]$.
Only the first term depends on $\varphi$'s and it is maximal 
for $\cos(\varphi_1 - \varphi_2) = \pm 1$. 
We put $\cos(\varphi_1 - \varphi_2) = -1$ and
prove that $L$ is maximal for $\theta_1 = \theta_2 \equiv \theta$ such that $\cos \theta = 1-p$
(the choice of $\cos(\varphi_1 - \varphi_2) = +1$ gives the same maxima).
For this choice $\frac{\partial L}{\partial \theta_1} = \frac{\partial L}{\partial \theta_2} = 0$,
i.e. the allowed $\theta$'s define stationary points.
To show that they correspond to a maximum
we compute second derivatives 
$\frac{\partial^2 L}{\partial \theta_1^2} = \frac{\partial^2 L}{\partial \theta_2^2}  = -p$
and $\frac{\partial^2 L}{\partial \theta_2 \partial \theta_1} = p (1-p)^2$
at the stationary points.
The derivatives $\frac{\partial^2 L}{\partial \theta_n^2}$ are negative
for all $p$, as it should be for a maximum.
It remains to show that the Hessian determinant is always positive.
In our case it is given by $-p^6+4p^5-6p^4+4p^3$
and indeed it is always strictly positive in the allowed range of $p$.
Using the optimal $\theta$'s the left-hand side equals $L = 2 p - 2 p^2 + p^3$.
The right-hand side of condition (\ref{WEIGHTED_CONDITION})
is given by $R = 2 p - 2 p^2 + 2 p^3$
and it is bigger than $L$ for all $p$.
Thus, the state (\ref{COLORED_NOISE}) is entangled
for any, no matter how small, admixture of $| \psi^- \rangle$.

\emph{Condition for density operators}.
All the steps in the proof of condition (\ref{WEIGHTED_CONDITION})
can be done without any reference to a specific representation of the state.
As a generalized scalar product in the operator space one just has to take a weighted-trace
with the positive semi-definite superoperator $G$, i.e.,
$(\rho_1,\rho_2)_G = \mathrm{Tr}(\rho_1 G \rho_2)$.
The sufficient condition for entanglement now reads:
if there is a positive superoperator $G$, such that
\begin{equation}
 \max_{\rho_{\mathrm{prod}}} \textrm{ } \mathrm{Tr}(\rho G
\rho_{\mathrm{prod}}) < \mathrm{Tr}(\rho G \rho),
\label{DENSITY_CONDITION}
\end{equation}
where we maximize over all pure product states $\rho_{\mathrm{prod}}$,
then state $\rho$ is entangled.

\emph{Bound entanglement}.
To show usefulness of condition (\ref{DENSITY_CONDITION}), consider a state of two four-level systems which is  bound entangled:
\begin{equation}
\label{BOUND1}
\rho_{\mathrm{B}} = \frac{\rho_0 + p \rho_+ + q \rho_-}{1+p+q}.
\end{equation}
The contributions to $\rho_{\mathrm{B}}$ represent one separable and two
entangled states. 
The separable contribution is
\begin{equation}
\rho_0 = \frac{1}{4} \left(|\! ++ \rangle \langle ++ \!| + |\! -- \rangle \langle -- \!| 
+ |2 \ 2 \rangle \langle 2 \ 2 | + | 3 \ 3 \rangle \langle 3 \ 3 | \right)
\end{equation}
with $| \pm \rangle = \frac{1}{\sqrt{2}}(|0\rangle \pm | 1 \rangle)$.
The two entangled contributions are
\begin{equation}
\rho_{\pm} = \frac{1}{4} \left( |\phi_{02}^{\pm} \rangle \langle \phi_{02}^{\pm} | +
|\phi_{12}^{\pm} \rangle \langle \phi_{12}^{\pm} | + |\phi_{03}^{\pm}
\rangle \langle \phi_{03}^{\pm} | + |\phi_{13}^{\pm} \rangle \langle
\phi_{13}^{\pm} | \right),
\end{equation}
where for $j=0,1$ we define
$|\phi_{j2}^{\pm} \rangle  =  \frac{1}{\sqrt{2}}(| j \ 2 \rangle \pm |2 \ j \rangle)$  
and 
$|\phi_{j3}^{\pm} \rangle  =  \frac{1}{\sqrt{2}}(| j \ 3 \rangle \pm |3 \ j' \rangle)$
with $j' = (j+1)\mod 2$.
For convenience, we introduce a real parameter $b$ defined by
$p = \frac{b+\sqrt{2}}{2}$ and $q = \frac{b-\sqrt{2}}{2}$ and consider $b \geq \sqrt{2}$.
The state $\rho_{\mathrm{B}}$ has a positive partial transposition for all $b \geq \sqrt{2}$.
In fact, it represents a partial transposition of the state generating a
classical probability distribution described by Renner and Wolf in their
search for bound information \cite{RENNER03}.
To reveal entanglement of $\rho_{\mathrm{B}}$, one can use our criterion
(\ref{DENSITY_CONDITION}) 
with $G$ representing projection on $\rho_+$,
$G = |\rho_+ )(\rho_+|$, where $| . )(. |$ denotes a
projector in the Hilbert-Schmidt space.
Since $\rho_+$ is orthogonal to both $\rho_0$ and $\rho_-$ and $\mathrm{Tr}\rho_+^2 = \frac{1}{4}$
the right-hand side reads
$(\rho_{\mathrm{B}}, \rho_{\mathrm{B}})_G = (\rho_{\mathrm{B}}, \rho_+)(\rho_+,\rho_{\mathrm{B}}) = \frac{p^2}{16(1+b)^2}$.
The left-hand side is given by
$\max(\rho_{\mathrm{B}},\rho_{\mathrm{prod}})_G = \max(\rho_{\mathrm{B}},\rho_+)(\rho_+,\rho_{\mathrm{prod}})
= \frac{p}{32(1+b)}$, where $\max(\rho_+,\rho_{\mathrm{prod}}) = \frac{1}{8}$.
A product state for which this maximum is achieved can be chosen among the products
which enter the definitions of $| \phi^{\pm} \rangle$'s.
Numerical results agree with $\frac{1}{8}$ being the maximum over the choice of any product state.
Thus, we identify entanglement whenever $\frac{b+\sqrt{2}}{b+1} >1$, i.e. for all $b \ge \sqrt{2}$.

\emph{Necessity of the criterion}.
The examples  suggest that the conditions 
are {\em necessary} for entanglement, i.e.
\begin{itemize}
\item
 for every entangled state there
exists a metric $G$ such that inequalities
(\ref{DENSITY_CONDITION}) or (\ref{WEIGHTED_CONDITION}) hold. 
\end{itemize}

Proof. The separable
states $\rho_{\mathrm{sep}}$ form a convex and compact set, $\mathcal{S}$,
in the  space of Hermitian operators, with a standard scalar product $(\varrho |
\varrho')=\mathrm{Tr}(\varrho \varrho')$. Any entangled state $\rho_{\mathrm{ent}}$ is at a non-zero distance from $\mathcal{S}$.
Consider two separable states $\rho_0$ and $\rho_1$, and let $\rho_0$ be the
separable state, which minimizes the distance to $\rho_{\mathrm{ent}}$. Due to convexity
of $\mathcal{S}$, any convex combination of these states, $\rho_{\lambda} =
(1-\lambda) \rho_0 + \lambda \rho_1$ is separable too. Thus,  the norm (length) of operator
$\gamma = \rho_{\mathrm{ent}} - \rho_{\lambda}$ is strictly positive, and not smaller than the norm of $\gamma_0 = \rho_{\mathrm{ent}} - \rho_0$: 
\begin{equation}
\label{H-S}
        \left\| \gamma \right\|^2 = \lambda^2 \| \rho_0 - \rho_1 \|^2 + 2
\lambda (\rho_0 - \rho_1 | \gamma_0) + \| \gamma_0 \|^2\geq \left\| \gamma_0 \right\|^2.
\end{equation}
%%%%%%%%%%%%%%
The inequality can be saturated only for  $\lambda = 0$, otherwise
it contradicts our assumptions. Thus, the derivative of the left-hand side with
respect to $\lambda$, at $\lambda=0$, cannot be negative. This requires that
$(\rho_0 | \gamma_0) \geq (\rho_1 | \gamma_0)$ for all separable
$\rho_1$.    
We also have $\| \gamma_0 \|^2 =
(\rho_{\mathrm{ent}} - \rho_0 | \gamma_0) > 0$.
Consequently, for all separable states
$\rho_{\mathrm{sep}}$, one has
\begin{equation}
\label{NECESSITY}
(\rho_{\mathrm{ent}} | \gamma_0) > (\rho_{\mathrm{sep}} |
\gamma_0).
\end{equation}

Moreover, the definition of $\gamma_0$ implies that $(\rho_{\mathrm{ent}} | \gamma_0) > 0$. To show this, we decompose $\rho_{\mathrm{ent}}$ and $\rho_0$ into $\frac{1}{d} \openone + X$ and $\frac{1}{d} \openone + Y_0$, respectively. Convexity of $\mathcal{S}$, together with the fact that the maximally mixed state $\frac{1}{d} \openone$ is separable, require that $\rho_0^{\lambda} = \frac{1}{d} \openone + \lambda Y_0$ is a separable state for all $0 \leq \lambda \leq 1$. In this range the lenght of $\delta= \rho_{\mathrm{ent}} - \rho_{0}^{\lambda} = X - \lambda Y_0$ reads
\begin{equation}
\label{NORM_GAMMA}
\left\| \delta \right\|^2 = \left\| X \right\|^2 - 2 \lambda (X|Y_0) + \lambda^2 \left\| Y_0 \right\|^2 \geq \left\| \gamma_0 \right\|^2 > 0
\end{equation}
For (\ref{NORM_GAMMA}) to hold, one needs $\frac{d}{d \lambda} (\left\| \delta \right\|^2) \leq 0$ at $\lambda=1$. This requires that $(X|Y_0) \geq \left\| Y_0 \right\|^2$. Since $X-Y_0 = \gamma_0$ is a traceless operator one has $(X-Y_0 | Y_0) = (\gamma_0 | \rho_{0})$. Thus the last inequality is equivalent to $(\gamma_0 | \rho_{0}) \geq 0$. By combining this fact with inequality (\ref{NECESSITY}) and the separability of $\rho_0$, we arrive at the strict positivity of $(\rho_{\mathrm{ent}} | \gamma_0)$. This allows us to multiply both sides of (\ref{NECESSITY}) by $(\rho_{\mathrm{ent}} | \gamma_0)$  without reversing the inequality's direction. It results in 
$(\rho_{\mathrm{ent}} | \gamma_0)(\gamma_0|\rho_{\mathrm{ent}}) >
(\rho_{\mathrm{sep}} | \gamma_0)(\gamma_0|\rho_{\mathrm{ent}})$
and we conclude that, for every entangled state $\rho_{\mathrm{ent}}$ one can find 
a positive semi-definite superoperator $G_{\gamma_0}$, the action of which can be symbolically expressed as $G_{\gamma_0} = |
\gamma_0 \left)\right( \gamma_0|$, such that for all separable states $\rho_{\mathrm{sep}}$
\begin{equation}
 (\rho_{\mathrm{sep}} | \ G_{\gamma_0}\ | \rho_{\mathrm{ent}}) <
(\rho_{\mathrm{ent}}| \ G_{\gamma_0}\ | \rho_{\mathrm{ent}}).
\end{equation}
Since this inequality is valid for all separable states
$\rho_{\mathrm{sep}}$, it is also valid for all pure product states
$\rho_{\mathrm{prod}}$. QED.

\emph{Relation with the entanglement witnesses}.
The Hermitian operator $\gamma_0$ of the above proof is related to the entanglement witness $W$ identifying entanglement of $\rho_{\mathrm{ent}}$. Simply, one has $W = q \openone - \gamma_0$, where $q =
\max_{\rho_{\mathrm{sep}} \in \mathcal{S}} (\rho_{\mathrm{sep}}| \gamma_0)$. Indeed,  $(W | \rho_{\mathrm{sep}}) \geq 0$ for all
separable states $\rho_{\mathrm{sep}}$ and $(W | \rho_{\mathrm{ent}})<0$.
Conversely, each entanglement witness $W$ defines a hermitian operator
$\gamma_0 = w \openone - W$, where $w = \max_{\rho_{\mathrm{sep}} \in \mathcal{S}} (W |
\rho_{\mathrm{sep}})$. However, one should stress 
that many entanglement identifiers $G$ {\em do not} have their witness counterparts. 
If $G$ is not a projector, there is no entanglement witness corresponding to it.
In particular, {\em only one} of the identifiers of our examples
can be associated with an entanglement witness!

\emph{Summary}.
We have derived simple sufficient conditions for entanglement of distributed quantum states.
Their generalization gives a necessary and sufficient separability criterion. 
The set of entanglement identifiers defined by our
criterion is strictly richer than the set of entanglement witnesses.
Moreover, the discussed examples indicate that
many identifiers not corresponding to any standard entanglement witness
are particularly useful.

\emph{Acknowledgments}.
W.L. is supported by a Stipend of the Foundation for Polish Science. 
We acknowledge support of the Austrian Science Foundation FWF (No. P19570-N16), 
the European Commission, Project QAP (No. 015846 and 015848),
and the FWF Doctoral Program CoQuS.
The collaboration is a part of an \"OAD/MNiSW program.

\end{document}